\documentclass[aps,pra,unsortedaddress, twocolumn,a4paper,amsmath,amssymb,showpacs,%
floatfix,]{revtex4}

\usepackage{graphicx}
\usepackage{epstopdf}

\usepackage[usenames,dvipsnames]{color}

\begin{document}
%
%
%
\title{Laser control strategies in full dimensional funneling dynamics: \\The case of pyrazine.}

\date{\today} 

\author{Samrit Mainali}
\affiliation{Universit\'e Paris-Saclay, CNRS, Institut des Sciences Mol\'eculaires d'Orsay, 91405 Orsay, France}
\author{Fabien Gatti}
\affiliation{Universit\'e Paris-Saclay, CNRS, Institut des Sciences Mol\'eculaires d'Orsay, 91405 Orsay, France}
\author{Osman Atabek}\email{osman.atabek@universite-paris-saclay.fr}
\affiliation{Universit\'e Paris-Saclay, CNRS, Institut des Sciences Mol\'eculaires d'Orsay, 91405 Orsay, France}

\begin{abstract}
Motivated by the major role funneling dynamics plays in light-harvesting processes, we built some laser control strategies inspired from basic mechanisms such as interference and kicks, and apply them to the case of pyrazine. We are studying the internal conversion between the two excited states, the highest and directly reachable from the initial ground state being considered as a donor, and the lowest as an acceptor. The ultimate control objective is the maximum population deposit in the otherwise dark acceptor, from a two-step process: radiative excitation of the donor, followed by a conical-intersection-mediated funneling towards the acceptor. The overall idea is to first obtain the control field parameters (individual pulses leading frequency and intensity, duration and inter-pulse time delay) for tractable reduced dimensional models basically describing the conical intersection branching space. Once these parameters are optimized, they are fixed and used in full dimensional dynamics describing the electronic population transfer. In the case of pyrazine, the reduced model is 4 dimensional, whereas the full dynamics involve 24  vibrational modes. Within experimentally achievable electromagnetic field requirements, we obtain a robust control with about $60\%$ of the ground state population deposited in the acceptor state, while about $16\%$ remains in the donor. Moreover, we anticipate a possible transposition to the control of even larger molecular systems, for which only a small number of normal modes are active, among all the others acting as spectators in the dynamics.
\end{abstract}


\maketitle




\section{Introduction}
\label{sec:intro}
Funneling dynamics is a key mechanism in the framework of artificial light-harvesting processes, biological antennas, or organic photovoltaic devices  \cite{Scholes_2017, Chin_2013, Tanimura}. Transfer from an initial donor to a target acceptor proceeds through an excited state gradient, leading to a final localization from which the photonic energy may be captured. In complex molecular systems, such energy transport mechanisms are driven by ultra-fast non-adiabatic transitions through conical intersections (CI) \cite{Thorwart, Arnold, Mangaud, Koeppel, Yarkony, Yarkony2000}. Investigation of the coherent preparation of the donor state from the initial ground state and the control of the subsequent dynamics  using intense, ultra-short laser pulses are major issues for modifying the energy transport mechanisms \cite{Roscioli, Hu, Tomasi}. In this context, we have recently studied a rather complex molecular system, namely polyphenylene ethynylene dendrimer, with a control objective aiming at the coherent preparation of two donor states involved in the dynamics, or their symmetric versus asymmetric superposition \cite{JCP_155}. In these systems, the acceptor state is radiatively coupled to the ground state.  It is only by discarding this direct excitation that we could build control strategies to increase the coherence (asymmetry) lifetime to populate the acceptor through the vibrational baths of the donors.

In the present work, we address the coherent control of the pyrazine (C$_4$H$_4$N$_2$) photophysics and more precisely its $S_2\rightarrow S_1$ internal conversion leading to challenging issues, both with respect to its ultra-fast non-adiabatic CI transfer (internal conversion time-scale of 20 fs \cite{Stert, Raab, Seidner, Worth1996, Woywood}) and to its 24-dimensional vibrational excitation dynamics. Some previous works in the literature are also concerned by the coherent control of population transfer from $S_0$ to $S_2$ and $S_1$ electronic states putting the emphasis on a significant delay of the internal conversion (or, in general, radiationless transition), which is considered as an undesired phenomenon \cite{Brumer_2012, Seideman_PRL, Seideman_PRA}. This is in relation with the fact that a successfully populated excited state ($S_2$) would be readily available for applications (such as observation of desired photoproducts, mode selective chemistry, light-triggered molecular rotors, separation of racemic mixtures) for only a brief period of time. At this respect, refs.\cite{Seideman_PRL, Seideman_PRA} focus specifically on an optimal control scheme to suppress radiationless transitions on ns time scales after the external control is over. As for ref.\cite{Brumer_2012}, the authors proceed to a detailed study of the $S_0 \rightarrow S_2 \rightarrow S_1$ process putting the emphasis on maximal $S_2$ population during the control scheme (about 50 fs duration). Still other works \cite{Christopher_2005, Christopher_2006} refer to more general schemes by using overlapping resonances. The main computational limitation is identified as the reduced number of vibrational modes, usually 4 rather than 24. However, within some approximations, model extensions to 24 dimensions have been achieved, making use of a so-called QP partitioning \cite{Christopher_3}. Motivated by light-harvesting systems, we pursue an opposite goal. Our purpose is to maximize the population of the energetically lower and thus more stable excited state $S_1$. The basic idea is to deposit the exciton energy from the bright donor state $S_2$ to the dark acceptor state $S_1$, through a funneling mechanism triggered by a conical intersection and non-adiabatic internal coupling, taking advantage of a decreasing electronic gradient from $S_2$ to $S_1$. 

Control objectives being opposite, the  strategies we are developing also differ completely from the ones referred to in previous works. More precisely, when maximizing $S_2$ population, the emphasis is put on stable, electronically localized eigenstates among the very dense manifold of vibronically coupled levels. Refs.\cite{Seideman_PRL, Seideman_PRA} are concerned by an optically coherent superposition of a specific eigenstate localized in the $S_2$ potential well. Ref.\cite{Brumer_2012} also deals with high-lying $S_2$ vibrational states with significantly slower decay rates into $S_1$. A complete study of excitation frequencies is undertaken to resonantly reach superposition of such excited vibrational states, using rather long duration pulses (180 fs for \cite{Seideman_PRL, Seideman_PRA} and 40 to 100 fs for \cite{Brumer_2012}). Some other works deal either with the dynamic Stark effect used as a basic mechanism to shift the CI away from the Franck-Condon region \cite{Sala, Saab, Sala2015}, or with optimal control theory \cite{Wang}, but they are limited to three or four modes vibronic couplings. We are adopting different control strategies based on two mechanisms: either  interference, or sudden excitation kicks, as extended to high dimensional complex molecular systems. This is conducted by using ultra-short broad-band pulses, without a specific resonant excitation frequency. 
We note that such strategies remain in the spirit of what has been suggested in the preliminary work of Ref.\cite{Ferretti}.
Our control schemes  are built so as to decrease the ground state $S_0$ population, and simultaneously reach the optimal contrast favoring $S_1$ population as compared to the one of $S_2$. As a signature of stable deposit of electronic energy, we are interested in post-pulse $S_1$ population, taking into account the dephasing processes of the full 24-modes dynamical description of pyrazine.
On the experimental side, it is worth mentioning that several works are addressing the control of the internal conversion to the energy transfer ratio, using in particular, open learning loop setups and evolutionary algorithms \cite{Savolainen, Herek, Hauer, Buckup}. In particular, the multi-pulse excitation (a degenerate four-wave-mixing sequence) has its counterpart in the present work, in terms of ultrashort pulse trains to propel the wave packet towards the conical intersection.

The manuscript is organized as follows. Section II is devoted to the Frenkel vibronic Hamiltonian involving the 24-modes model of pyrazine, with all relevant parameters taken from \cite{Worth}. The methodology adopted to treat this model is Multi Configurational Time Dependent Hartree (MCTDH) \cite{Meyer:book, mctdh:package, Meyer, Worth, Manthe, Beck} which is shortly reviewed with respect to its numerical convergence criteria in the CI branching space. The post-pulse field-free dynamics and its recursive patterns in terms of wavepacket vibrational periods on both $S_2$ and $S_1$ potential energy curves are analyzed. A detailed interpretation is provided when going from 2D to 4D, and 24D models. Control strategies are presented in Section III, putting the emphasis on control observable and parameters on the one hand, and interference and kick mechanisms from weak to strong field regimes, on the other hand.   Results are presented in Section IV, for both the mechanisms with their interpretation. Conclusions and some perspectives are given in Section V.


\section{Methodology: Model Hamiltonian and numerical techniques.}\label{sec2}

\begin{figure} 
	\includegraphics[angle=-90, width=0.9\linewidth]{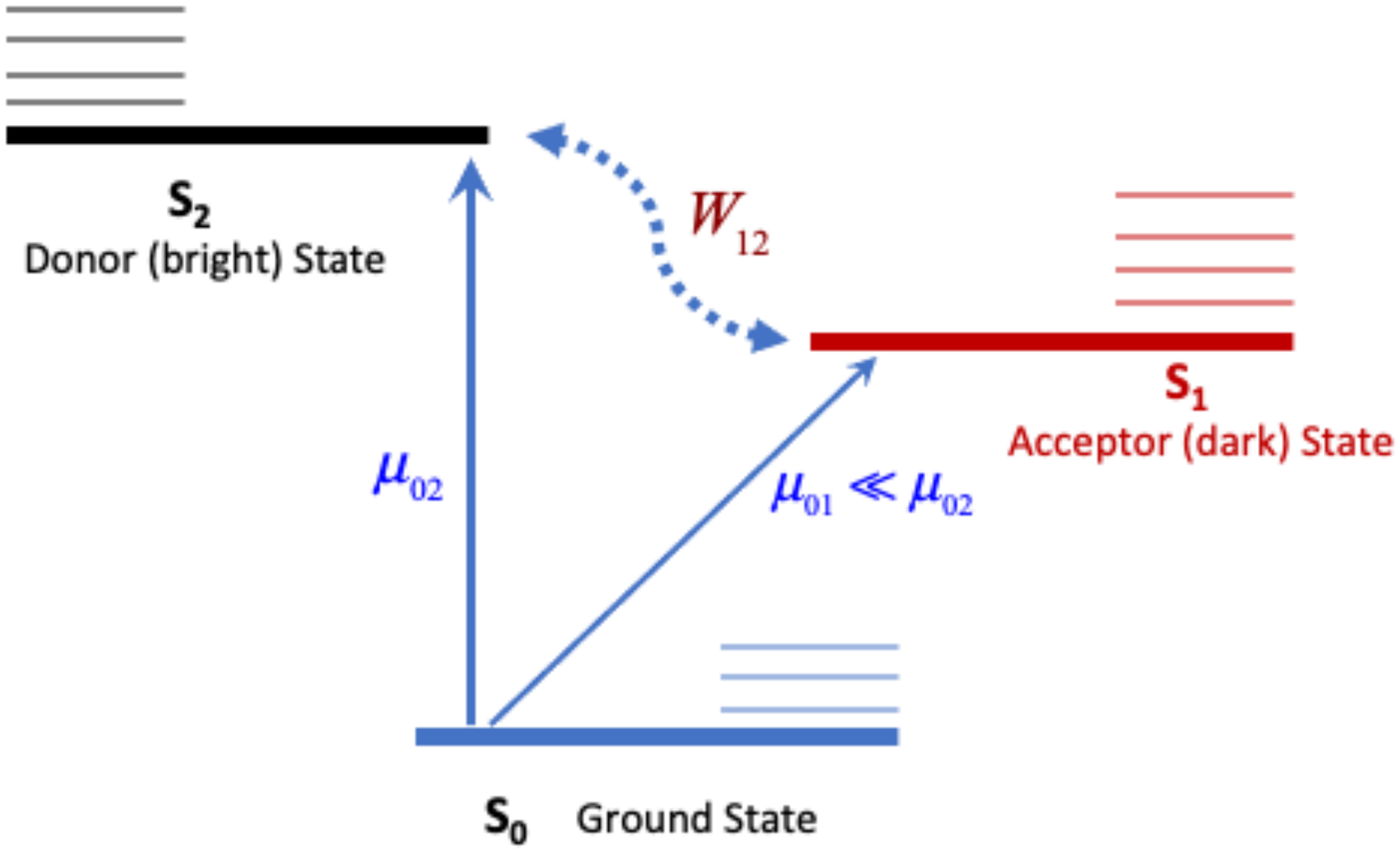}
	\caption{Schematic view of the three-state model used for pyrazine. The ground state $S_0$ is indicated in thick solid blue line. The bright donor state $S_2$ is in thick solid black line, and the dark acceptor state $S_1$ is in thick solid red line. The thin horizontal lines are for the corresponding vibrational states. Are also indicated the interstate coupling $W_{12}$ in dotted blue line, and the transition dipole moments $\mu_{01}$ and $\mu_{02}$. }
	\label{schematic}
\end{figure}

As in many previous works, we are considering three electronic states of pyrazine, namely the ground state $S_0(^1A_g)$, together with the two lowest excited  electronic states, $S_1(^1B_{3u})$ and $S_2(^1B_{2u})$. These states are well-separated from others, close in energy, and vibronically coupled to each other \cite{Raab}. Figure (\ref{schematic}) displays a schematic view of the corresponding states and their couplings. 

Following the usual model of displaced harmonic oscillators  in open quantum systems, we work with a linear vibronic Hamiltonian expressed in a diabatic representation. In this Section, we examine the field-free vibronic Frenkel Hamiltonian, the MCTDH method retained for solving it, and the generic field-free dynamics after launching part of the ground state wavepacket on the excited states. That is more precisely conducted with successively 2D, 4D and 24D models for a clear understanding of the role of dimensionality.


\subsection{Vibronic Hamiltonian} \label{Hamiltonian}
We briefly recall the generic field-free Frenkel  Hamiltonian which reads \cite{Koeppel1984}
\begin{equation}
H_S(Q)=\sum_n\vert n \rangle H_{nn}(Q) \langle n \vert + \sum_{n \neq m}\vert n\rangle H_{nm}(Q) \langle m \vert
\label{HS}
\end{equation}
where $n$ and $m=0,1,2$ denote the electronic states and Q collectively the nuclear coordinates.
Using the linear vibronic coupling of ref. \cite{Worth} in mass weighted coordinates, we have:
\begin{equation}
H_{nn}(Q)=\tilde{\epsilon}^{(n)}+\sum_{\alpha}^{N_{\alpha}}\{P_{\alpha}^2 + {\omega_{\alpha}^{(n)}}^2(Q_{\alpha}-d_{\alpha}^{(n)})^2 \}
\label{Hnn}
\end{equation}
and 
\begin{equation}
H_{nm}(Q)=\sum_{\alpha}^{N_{\alpha}}\lambda _{\alpha}^{(nm)}Q_{\alpha}
\label{Hnm}	
\end{equation}
where $Q_{\alpha}$ and $P_{\alpha}=-i \hbar \frac{\partial}{\partial Q_{\alpha}}$ are the coordinate and  momentum associated with the mode $\alpha$. There are $N_{\alpha}$ such modes in the electronic state $n$ with frequency $\omega_{\alpha}^{(n)}$ and displacement $d_{\alpha}^{(n)}$.
Reorganizing Eq.({\ref{Hnn}}), one gets:
\begin{equation}
H_{nn}(Q)={\epsilon}^{(n)}+\sum_{\alpha}^{N_{\alpha}}\{P_{\alpha}^2 + {\omega_{\alpha}^{(n)}}^2 Q_{\alpha}^2\} +\sum_{\alpha}^{N_{\alpha}}\kappa_{\alpha}^{(n)} Q_{\alpha} 
\label{Hnnr}
\end{equation}
where the gradient at the reference point ($Q_{\alpha}=0$)  is given by: $\kappa_{\alpha}^{(n)}=-2 {\omega_{\alpha}^{(n)}}^2d_{\alpha}^{(n)}$, and the gradient of the interstate coupling is noted as: $\lambda _{\alpha}^{(nm)}=W_{12}$. Finally, the renormalized site energy is: ${\epsilon}^{(n)}=\tilde{\epsilon}^{(n)}+{\omega_{\alpha}^{(n)}}^2{d_{\alpha}^{(n)}}^2$.
Table I of ref. \cite{Worth} collects all the parameters that are relevant for the system Hamiltonian given by Eq.(\ref{HS}). We have to position $S_2$ with respect to $S_0$, and for this we take the value 4.84$eV$ from \cite{Schneider, Woywood}. 

It is to be noted that the two normal modes that span the conical intersection branching space, using the symmetry group notations,  are $Q_{6a}$ for state $S_1$ and $Q_{10a}$ for state $S_2$ \cite{Koeppel1984}.
As mentioned before, the three models will be referred to as 2D ($N_{\alpha}=2$) involving $Q_{6a},Q _{10a}$, 4D ($N_{\alpha}=4$) when enlarging the branching space to $Q_{9a}$ and $Q _{1}$, and the full 24D ($N_{\alpha}=24$) description involving all normal modes.



\subsection{MCTDH survey} \label{MCTDH}
The MCTDH approach for multi electronic states and multi-mode nuclear dynamics has been presented in detail in the literature. For pyrazine, with its three electronic states in consideration, the multistate vibrational wave function is taken as a 3-dimensional column vector $(\Psi_1, \Psi_2, \Psi_3)^T$, of single-state nuclear wave functions $\Psi_n, (n=1,2,3)$, $T$ being the transpose. The total electro-nuclear eigenvector is then written as:
\begin{equation}
\vert \Psi^{Tot}(Q,t)\rangle=\sum_{n=1}^3\Psi_n(Q,t)\vert n\rangle
\label{mcs}
\end{equation}
$n$ labels the electronic states and the unknown nuclear wave functions are solutions of the following close-coupled equations:
\begin{equation}
-i\hbar \frac{\partial}{\partial t}\Psi_n(Q,t)=H_{nn}\Psi_n(Q,t) + \sum_{m\neq n}H_{nm}\Psi_m(Q,t)
\label{cc}
\end{equation}

A standard multiconfiguration approach consists in expanding these functions on a time-independent basis set $\Phi_{J,n}$, with time-dependent coefficients $A_{J,n}$, as:
\begin{equation}
\Psi_n(Q,t)=\sum_J^{N_J} A_{J,n}(t) \Phi_{J,n}(Q)
\label{total}
\end{equation}
The major improvement brought by MCTDH is that not only the coefficients but also the basis functions are taken as time-dependent. The challenge is that, by adapting the basis set functions to the temporal evolution, one must reduce the total number of these functions (i.e., $N_J$) for a given convergence criterion. To proceed along this line, the now time-dependent basis set functions $ \tilde \Phi_{J,n}(Q,t)$ are given as a tensorial product of time-dependent single-particle functions $\varphi_{j_{\alpha}}^{(\alpha)}(Q_{\alpha},t)$ describing a given normal mode $\alpha$. The complete expansion reads as:
\begin{equation}
\Psi_n(Q, t)= \sum_{j_1=1}^{n_1}...\sum_{j_f=1}^{n_f} A_{j_1,...j_f}(t) \prod_{j_\alpha}^{n_{\alpha}}\varphi_{j_{\alpha}}^{(\alpha)}(Q_{\alpha},t)
\label{mcMCTDH}
\end{equation}
The index $j_{\alpha}$ stands for one of the
 $n_{\alpha}$ possible single particle functions for the $\alpha$th mode. The number of configurations is thus given by the product $n_1...n_f$.
The single particle functions are ultimately expressed in a time-independent, so-called primitive basis set functions $\chi_{i_k}^{(\alpha)}(Q_k)$ as:
\begin{equation}
\varphi_{j_{\alpha}}^{(\alpha)}(Q_{\alpha},t)=\sum_{i_k}^{M_{\alpha}} c_{i_k}^{(\alpha, j_{\alpha})}(t)\chi_{i_k}^{(\alpha)}(Q_k)
\label{primitive_basis}
\end{equation}
 More precisely, for this primitive basis set, we have used 22 harmonic oscillator Discrete Variable Representation (DVR) functions for the modes $\alpha=6a, 10a$, whereas only 12 and 11 such functions are retained for the modes $\alpha=9a$ and $\alpha=1$ modes, respectively. All other modes, not playing a crucial part in the dynamics, are merely described by 4 DVR functions.  To further reduce the memory and the numerical effort, we proceed to another modification, by combining several degrees of freedom, as has previously been suggested \cite{Worth}. This mode combination scheme finally gives rise to 8 modes labeled $\tilde Q_j$.  ($j=1,...8$). In particular, $\tilde Q_1$ results from a linear combination of $Q_{6a}$ and $Q_{10a}$. For the others the combination scheme is indicated in parenthesis, i.e., 
 $\tilde{Q_2} (1, 9a, 8a)$, $\tilde{Q_3} (2, 4, 5)$, $\tilde{Q_4} (6b, 3, 8b)$, $\tilde{Q_5} (7b, 16a, 17a)$, $\tilde{Q_6} (12, 18a)$, $\tilde{Q_7} (19a, 13, 18b, 14)$ and $\tilde{Q_8} (19b, 20b, 16b, 11)$.
 \begin{table}
 	\setlength{\tabcolsep}{5pt}
 	\begin{tabular}{ccccccccc}
 		\hline
 		\hline
 		$n$ & $\tilde{Q_1}$ & $\tilde{Q_2}$ & $\tilde{Q_3}$ & $\tilde{Q_4}$ & $\tilde{Q_5}$ & $\tilde{Q_6}$ & $\tilde{Q_7}$ & $\tilde{Q_8}$\\
 		\hline\\
 		{$1$} & 10 & 10 & 4 & 4 & 4 & 4 & 4 & 4 \\[1mm]
 		{$2$} & 28 & 28 & 4 & 4 & 4 & 4 & 4 & 4 \\[1mm]
 		{$3$} & 28 & 28 & 4 & 4 & 4 & 4 & 4 & 4 \\
 		\hline
 		\hline
 	\end{tabular}
	 \caption[]{Number of single particle functions used for the combined modes (24D). The size of the DVR for each mode is given in the text.}
	\label{BasisD Size:24}
 \end{table}
 Numerical convergence is obtained by a much smaller basis set of single particle functions which are no more one-dimensional. 
 
 The equations of motion for the coefficients $A$ of Eq.(\ref{mcMCTDH}) result from the time-dependent evolution equation involving the  total Hamiltonian, i.e. the one of the system, together with the molecule-field coupling. The close coupled system of differential equations (Eq.(\ref{cc})) is solved by projecting on the basis function of the combined modes. The initial condition, at time $t=0$, being taken as 1 for the vibrationless ground state $n=1$, and 0 for all other states, the nuclear wave functions are built following Eq.(\ref{mcMCTDH}). Finally the time-dependent population in each electronic state is given as:
\begin{equation}
P_n(t)=\int \vert \Psi_n(Q,t)\vert ^2  dQ
\label{population}
\end{equation}


\subsection{Field-free dynamics} 
\begin{figure} 
	\includegraphics[angle=-90, width=0.9\linewidth]{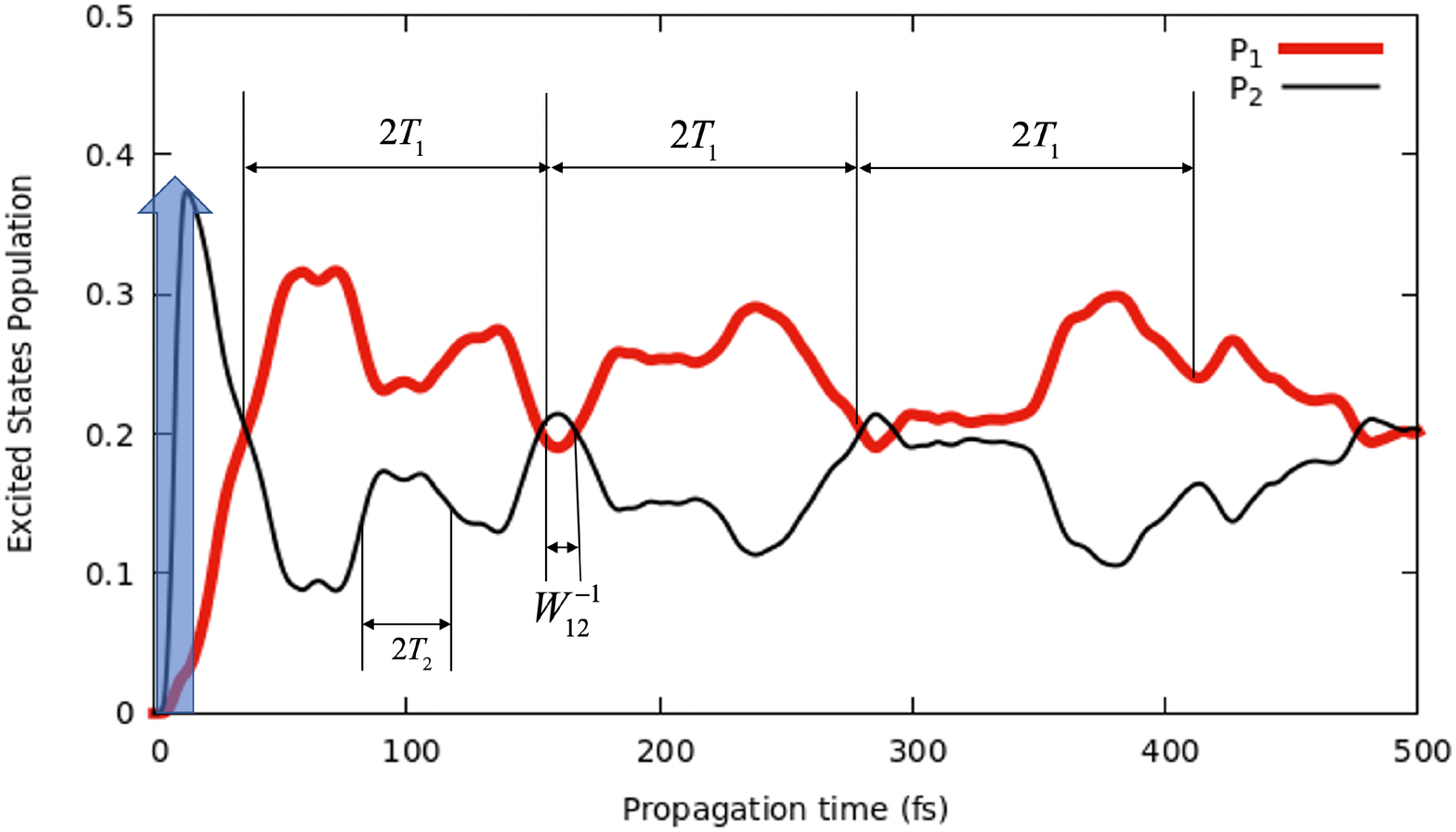}
	\caption{Excited states populations as a function of propagation time in a minimal 2D model involving the CI branching space. $P_2$ and $P_1$ are respectively indicated by thin black and thick red solid lines. The blue vertical arrow at t=0 symbolizes the vertical launching of the $v=0$ wavepacket from the initial state $S_0$. $T_1$ and $T_2$ are respectively, the vibrational periods of the excited states $S_1$ and $S_2$. $W_{12}^{-1}$ is a notation for the CI characteristic transfer time.}
\label{ff_dynamics}
\end{figure}

In order to have a typical generic overview of the $S_2 \rightarrow S_1$ population transfer, we now proceed to three dynamical calculations extending over a period of 500 fs. More precisely, we successively analyze 2D and 4D approximate models and relate them with the full 24D model. The minimal 2D model merely involves the two normal modes $\alpha= 6a $ and $\alpha= 10a$ building up the conical intersection branching space. The population transfer dynamics is induced by some Franck-Condon vertical launching of the initial vibrationless ($v=0$) $S_0$ wavepacket on the excited states. To fix the ideas, this is practically done referring to a low intensity ($5 \times  10^{12}W/cm^2$), short  duration (14 fs), resonant laser pulse, which is symbolized by the vertical blue arrow of figure (\ref{ff_dynamics}) positioned at time t=0. It is worthwhile noting that the precise parameters of this excitation are, but an illustrative example for the kind of sudden excitation we refer to initiate the transfer dynamics. During the 14fs $S_0 \rightarrow S_2$ resonant excitation process, it is basically the $S_2$ state which is populated up to $P_2=0.37$, whereas $P_1$ does not exceed 0.03. The short pulse here is just to initiate a typical population transfer from the initial ground $S_0$ to the bright donor state $S_2$, without any control purpose. Actually, what we call free-field dynamics is to be understood as the wave packet dynamics following the switch-off of this pulse.

The post-pulse field-free dynamics starts with this configuration, and proceeds towards the descending energy gradient from $S_2$ to $S_1$ electronic potential energy surfaces coupled by $W_{12}$. In particular, due to the conical intersection, an important amount of population is transferred to $S_1$. After a delay corresponding to the $S_1$ vibrational period ($T_1$=56fs with the parameters of our model), the wavepacket reflecting on the outer right turning point returns back to the CI region with a partial back transfer to $S_2$, and then evolves with another oscillation with the vibrational period ($T_2$=44 fs) of the state $S_2$. The combination of these two oscillation modes in the vibrational baths of $S_1$ and $S_2$ gives rise to a recursive pattern in the population dynamics which is then periodically repeated. Figure (\ref{ff_dynamics}) provides a complete illustration  of such recursive patterns. Calculations based on longer propagation times show that the revival structures last for more than 1ps. They are progressively attenuated and their periodicity is slightly altered due to wavepacket broadening.
\begin{figure}
		\centering
	\includegraphics[angle=-90, width=1.0 \linewidth]{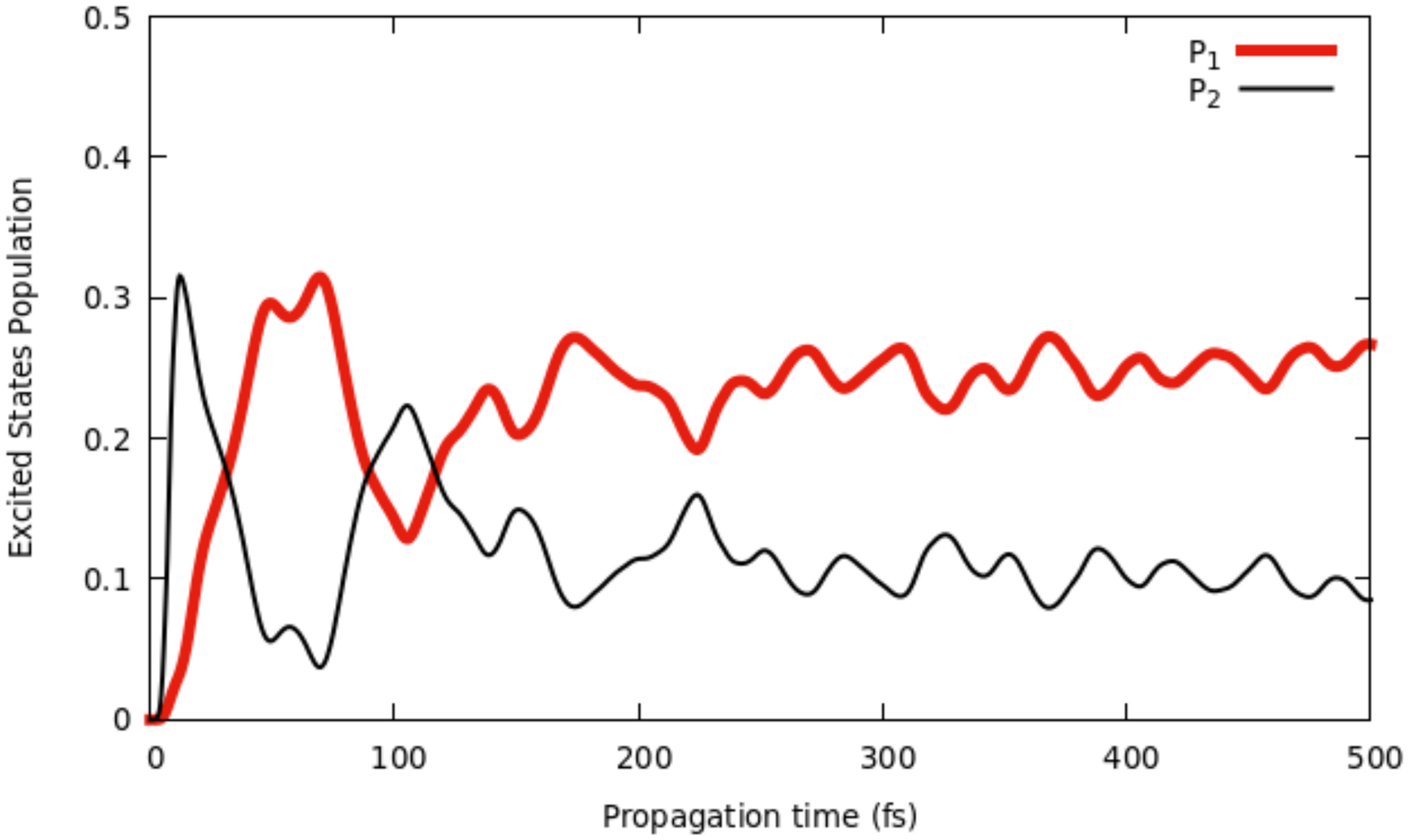}
	\centering
	\includegraphics[angle=-90, width=1.0 \linewidth]{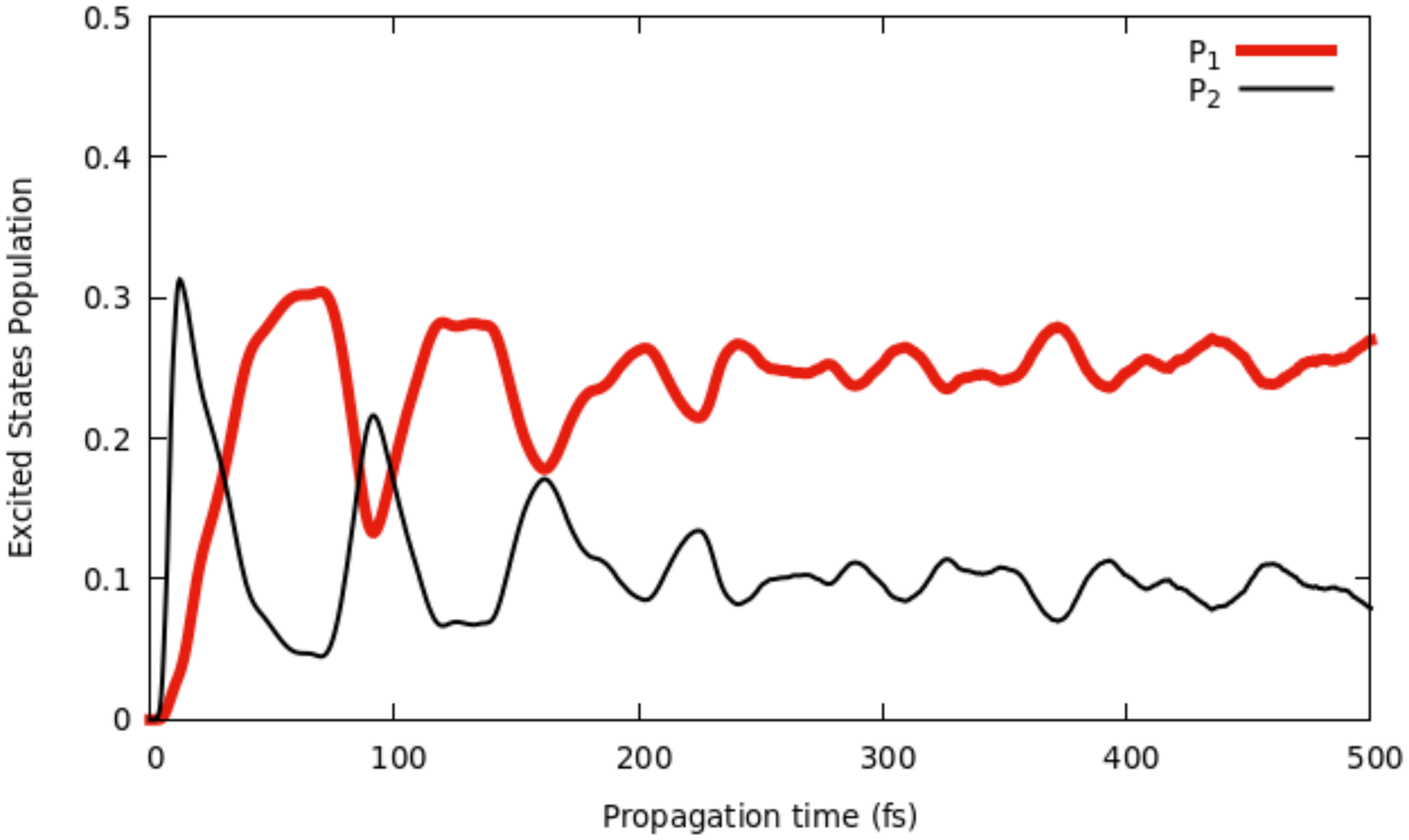}
	\caption{Excited states populations as a function of propagation time in 4D (upper panel) and full 24D models (lower panel). Same notations as for figure (\ref{ff_dynamics})}
	\label{ff_4and24}
\end{figure}
We now proceed by including higher degrees of freedom through additional normal modes. Figure (\ref{ff_4and24}) displays such calculations over 500 fs of propagation time, with the same excitation initiating the population transfer from $S_0$. It is to be noted that the additional degrees of freedom in multidimensional calculations dramatically increase the density of vibrational states. As a consequence, the specificity of the $T_1$ and $T_2$ vibrational periods of the $\alpha=6a$ and $\alpha=10a$ normal modes, building up the 2D model, is partially lost. The wavepacket is spreading over all the additional modes through all the vibrational states which form a quasi-continuum. Typically after 250 fs, the revival structures are completely suppressed. The $P_1$ and $P_2$ populations are moderately oscillating and progressively stabilizing in time.

\section{Coherent control.}\label{sec3}

Pyrazine molecule is studied in planar geometry and assumed to be oriented in a plane Oyz orthogonal to the propagation direction Ox of the electromagnetic field $E(t)$. The laser is supposed to be linearly polarized along the Oz axis. The time-dependent total Hamiltonian is written in the length gauge and within the dipole approximation as:
\begin{equation}
H(Q;t)=H_S(Q) + V(Q;t)
\label{Htot}
\end{equation}
$H_S$ being the molecular Hamiltonian taken from Eq.(\ref{HS}) and $V$ the radiative coupling:
\begin{equation}
V(Q;t)=-\mu(Q) E(t)
\label{Vrad}
\end{equation}
The transition dipole matrix elements between the ground $S_0$ and excited $S_1$, $S_2$ states are respectively noted $\mu_{01 }$ and $\mu_{02 }$. Their explicit spatial expansions in terms of the normal mode coordinates are given by \cite{Sala}:
\begin{equation}
\mu_{01}(Q) = \xi_{10a}^{(01)}Q_{10a}
\label{mu1}
\end{equation}
\begin{equation}
\mu_{02}(Q) = \mu_{02}(0) + \sum_{\alpha\neq 10a}\xi_{\alpha}^{(02)}Q_{\alpha} + \frac{1}{2}\rho_{10a}^{(02)}Q_{10a}^{2}
\label{mu2}
\end{equation}
It is worthwhile noting that for symmetry arguments, $\mu_{02 }$ is much larger than $\mu_{01 }$ due to its permanent dipole component $\mu_{02}(0)$. Numerical values we are using are extracted from ref.\cite{Sala2015}. It is also to be noted that we are working in such field conditions that referring to the polarizability is not necessary, as opposite to Ref. \cite{Sala2015}, concerned by very strong intensities and excitation conditions far from resonance.
We will now examine the relevance of some control observable, together with two strategies  to reach them in an optimal way, by exploiting two different mechanisms; namely, interference and kicks.

\subsection{Control observable.}

The primary objective of this study is to optimize the population $P_1$ of the acceptor state $S_1$. But for pyrazine, the acceptor state $S_1$ is a quasi-dark one, as is clear from Eq.(\ref{mu1}), practically without the possibility of a direct radiative excitation from the ground state. The population transfer process first proceeds with the radiative excitation of the ground $S_0$ state to the bright excited $S_2$, which then acts as a donor. The second step is monitored by the interstate coupling $W_{12}$ leading to a CI-induced strong non-adiabatic transfer from $S_2$ to $S_1$. 
A rather intuitive control objective could be taken as the maximization of the ratio $r(t)=P_1(t)/P_2(t)$. But this goes with the difficulty of getting very high ratio $r$, between actually very small  excited states populations, that is without any practical interest. To properly take into account the two steps, excitation and funneling, we built a population contrast type control observable combining two requirements: (i) maximization of the excited states populations, which actually amounts the minimization of the residual population $P_0$ of the ground state $S_0$, and (ii) among the two excited states, bring the maximum of population on the dark acceptor state $S_1$, or in other words, increase the contrast between excited states population. Such a time-dependent contrast is given by:
\begin{equation}
\tilde{p}(t)= \frac{P_1(t)-P_2(t)}{P_0(t)}
\label{contrast_t}
\end{equation}

It is worth noting that, in terms of $r$ Eq.\ref{contrast_t} can be written as:
\begin{equation}
\tilde{p}(t)= \frac{r(t)-1}{1/P_2(t)-[r(t)+1]}
\label{contrast_r}
\end{equation}
Maximizing $\tilde{p}$ requires simultaneously maximizing the numerator, that is $r>1$, and minimizing the (always positive) denominator, which is achieved with $P_2(t)<1/[1+r(t)]$. Finally, combining these two requirements, in particular limiting the $P_2$ increase, our control objective goes much beyond the passive observation of the acceptor state  population $P_1$ increase (i.e., increasing the donor state population $P_2$ and waiting for its CI mediated transfer towards the acceptor).

On numerical grounds, in weak fields, $P_0$ would be considered not less than 0.5, meaning that no more than half of the initial population is assumed to be transferred to the excited states. As a consequence, even in the case of the best contrast that could ideally be reached between the donor and acceptor states (that is $P_1=0.5$ and $P_2=0$), the maximum possible value for $\tilde{p}$ should not exceed 1. In other words, it is only in the strong field regime that we could expect a contrast $\tilde{p} > 1$.

Finally, instead of the time-dependent contrast given by Eq.(\ref{contrast_t}), we choose a scalar control observable by taking time averages of $\tilde{p}(t)$. 
\begin{equation}
p = \frac{1}{[t_{max}-t_{min}]} \int_{t_{min}}^{t_{max}}  \tilde{p}(t)dt
\label{contrast}
\end{equation}
This involves the definition of time windows over which the average is performed. A local contrast can be defined over a time-window displaying the maximum of contrast, which happens during the first vibrational pattern, lasting over 45 fs, as illustrated in figure \ref{ff_dynamics}, and leading to the choice $[t_{min},t_{max}]=[45 fs, 90 fs]$. Although during this 45 fs time-window some experimental observable are reachable, more interestingly, an asymptotic contrast is adopted in the following defining a semi-finite time interval above 300 fs. Incidentally, it is to be noticed that 300 fs corresponds to typical coherence times for such molecular systems and an upper limit 500 fs is taken as our final propagation time, leading to $[t_{min},t_{max}]=[300 fs, 500 fs]$

\subsection{Control fields.}
\begin{figure} 
	\includegraphics[angle=-90, width=0.9\linewidth]{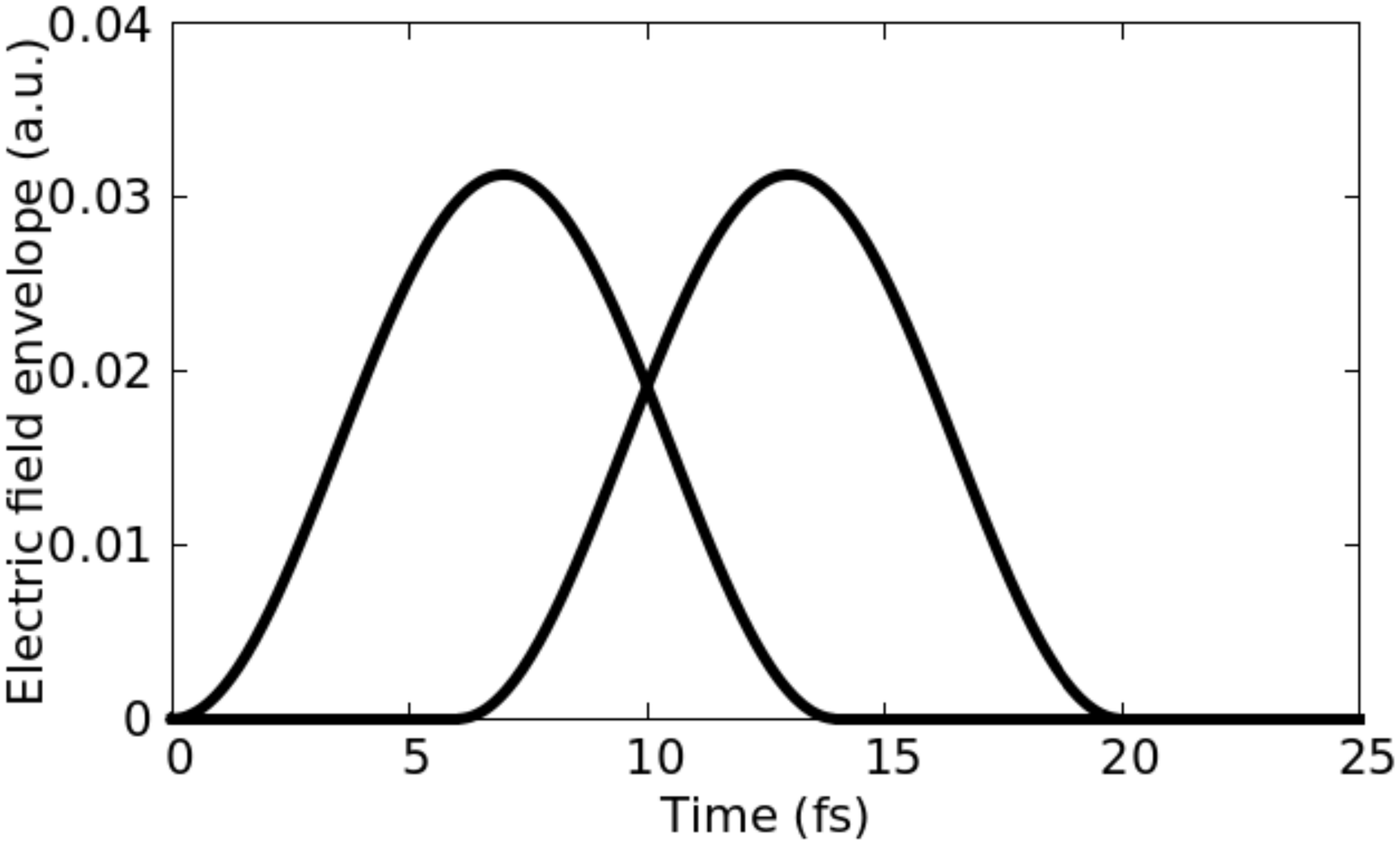}
	\caption{Sine-square laser electric field envelopes for two identical pulses as a function of time. The illustrated pulse duration corresponds to $T=14 fs$, whereas the inter-pulse time delay is $\tau=6fs$.}
	\label{envelope}
\end{figure}

In the following, we refer to two laser control strategies based on two mechanisms; namely interference and kicks. In both cases, the control field is made up of a train of $N$ time-delayed individual pulses. Two such pulses are illustrated in figure \ref{envelope}.
The general expression for the field involved in Eq.(\ref{Vrad}) is given by:
\begin{equation}
E(t)=\sum_{i=1}^{N} \sqrt{I} E_i(t) \sin(\omega t)
\label{pulse}
\end{equation}
where $I$ is the laser leading intensity and $\omega$ its frequency. $E_i(t)$ is the pulse envelope, taken as:
\begin{equation}
E_i(t)= \sin^2 [\frac{\pi}{T}(t-t_i)] H(t-t_i) H(t_i+T-\tau)
\label{pulse_envelope}
\end{equation}
where $T$ is the pulse duration. $H$ is the Heaviside function, being zero or one depending on whether its argument is negative or positive.
The time intervals $t_i$ are given as $t_i=(i-1)\tau$, where $\tau$ is the time delay between the two successive pulses. Moreover, we wish to disentangle the roles of strong field and interference effects. When referring to our control observable, a relevant comparison would require excitation conditions where the same total electromagnetic energy is deposited in the system by the pulse train over its full duration, and given by:  
\begin{equation}
\mathcal{A} = \int_0^{\infty} E^2(t) dt
\label{fluence}
\end{equation}
More precisely, we compare pulse trains with the same total energy, namely the one provided by the single pulse. To fix the ideas, for two identical pulses $E_1=E_2=\epsilon$, $\mathcal{A}$ can be written as: 
\begin{equation}
\mathcal{A} =  a \int_0^{\infty} \epsilon^2(t) dt
\label{fluence_2p}
\end{equation}
Referring to figure \ref{envelope}, three cases are examined: (i) $\tau \ge T$, no overlapping pulses, resulting in $a=2$; (ii) $\tau=0$, full overlapping, with $a=4$; and (iii) $0<\tau<T$, partial overlapping, leading to a delay-dependent $a$ ($2<a<4$), to be calculated according to Eqs. (\ref{fluence}, \ref{fluence_2p}). Our control fields amplitudes are normalized so as to provide the same total electromagnetic energy.  This is practically done by dividing the nominal intensities $I$ by $a$, or equivalently, the electric field amplitudes by $\sqrt{a}$.

We also wish to emphasize that when developing these strategies, we always have in mind the experimental feasibility with respect to pulse duration, intensities and maximum number of pulses in the train. For the interference scenario, the laser leading frequency $\omega$ is taken to be approximately resonant with the excited states vibrational levels at the Franck-Condon vertical region of the ground state. The optimal value which is adopted is $\omega=4.77 eV$, i.e., a wavelength of $\lambda=260 nm$ in the UV region. The pulse duration $T$ is calculated such that the corresponding energy band broadening covers enough vibrational levels taking part in the funneling process. For the kicks scenario, the resonance condition is no more a relevant requirement, since the ultra-short pulses involved in this strategy, would lead to excitation with even broader energy bands. All ultra-short pulses are taken identical and typical values which are retained ($T$=14 fs or even $T$=10 fs) correspond to band broadening ranging from  2382 cm$^{-1}$ up to 3335 cm$^{-1}$. Following a few attempts to roughly optimize the frequency $\omega$ and the duration $T$, we are finally left with two main control parameters, namely, the field intensity $I$ and the time delay $\tau$. 

\subsection{Coherent control strategies.}

The first strategy we are referring to relies on vibrational wave packets interference as a basic mechanism, envisioned either in a pump-probe or a pump-pump process. 
More precisely, as illustrated in figure \ref{schematic}, the pump-probe process involves two routes to reach $S_1$, starting from the ground state $S_0$. Route 1 (pump pulse) proceeds through the intermediate state $S_2$ with the lowest order transition amplitude given by \cite{Moiseyev}:
\begin{equation}
\mathcal{T}_1= \mathcal{V}+\mathcal{V}G_0\mathcal{V}
\label{vgv}
\end{equation}
where $G_0$ is the lowest order Green's function of the system Hamiltonian involving the diabatic electronic states, and $\mathcal{V}$ stands for interstate ($W_{12}$) and radiative ($\mu_{20}E_{1}$) couplings. 
More precisely, for Route 1, in the absence of direct radiative coupling $S_0 \rightarrow S_1$, one has:
\begin{equation}
\mathcal{T}_1= \langle S_1 \vert W_{12} \frac{1}{E_{S_0}-E_{S_2} +\hbar \omega} \mu_{20}E_1 \vert S_0\rangle 
\label{route1}
\end{equation}
\vspace {0.5cm}
\noindent 
where $E_{S_i}$ are the energies of the corresponding electronic 
states, and $E_1$ is the pump pulse electric field.
Route 2 (probe pulse) is the direct transition $S_0 \rightarrow S_1$, with a transition amplitude:
\begin{equation}
\mathcal{T}_2= \langle S_1 \vert  \mu_{01}E_2 \vert S_0\rangle 
\label{route2}
\end{equation}
with $E_2$, the probe pulse electric field.
When applying the two pulses with a time delay $\tau $, the pump-probe strategy finally results in an  $S_1$ population given by:
\begin{equation}
P_1 = \vert \mathcal{T}_1 +\mathcal{T}_2 e^{-i E \tau /\hbar}  \vert  ^2
\label{P1_pomp_probe}
\end{equation}
with $E=E_{S_0}+\hbar \omega$.
Due to very low transition dipole $\mu_{01}$ as  compared to $\mu_{02}$, the transition amplitude $\mathcal{T}_2$ is actually negligible unless very strong field amplitudes $E_2$ are used with possible ionization or dissociation damage on the molecule. This is why we rather focus on a pump-pump process retaining only route 1 and resulting in:
\begin{equation}
P_1 = \vert \mathcal{T}_1 +\mathcal{T}_1 e^{-i E \tau /\hbar}  \vert  ^2
\label{P1_pump_pump}
\end{equation}
Finally, the interference scheme we are considering through Eq.(\ref{P1_pump_pump}) is between the transition amplitudes $\mathcal{T}_1$, with a controlled delay $\tau$. Such interference mechanisms have already been recently exploited in the control of rotational anisotropy \cite{Chamaki}. A physical understanding of the mechanism can be obtained by considering the vibrational wavepacket back and forth oscillations in the excited states potentials. Following a first laser pulse leading to a vertical Franck-Condon launching from $S_0$ to $S_2$, the vibrational wavepacket oscillates during its early dynamics $t<70$ fs in the excited $S_2$ and $S_1$ states harmonic potentials. A second pulse is then applied with a controlled delay $\tau$ launching a second wavepacket from the ground state $S_0$, which can interfere with the first one, as they may overlap if the delay is adequately chosen. A constructive interference would produce an amplitude enhancement close to the CI, and thus an efficient population transfer from $S_2$ to $S_1$. Later on ($t>70$ fs), due to wavepacket spatial broadening and also its dispatching over additional degrees of freedom, the control efficiency is expected to be lost, as a consequence of less important successive wavepacket overlapping.

The interference mechanism can be extended and complemented by a kick mechanism, in the spirit of the one already discussed previously in the context of molecular alignment or orientation processes \cite{Sugny}. At this respect, we are considering a train of $N=5$ identical individual pulses in Eq.(\ref{pulse}). The kick mechanism assumes sudden excitation of the system. We are therefore addressing even shorter duration $T$=7 fs to 10 fs pulses without the specific need for a resonance condition. The molecule can be kicked every time when $P_1$ starts decreasing.

As has previously been stated, the control observable given by Eq.(\ref{contrast_t}), highly depends on the field intensity. In particular, it is only for strong field regimes that it can take the values exceeding 0.5. In order to fix the intensity regimes for the present system described within a 4D model, we proceed to dynamical calculations using both a single pulse or two (equal fluence) pulses with a typical delay of 13 fs, and for a series of increasing intensities. 
\begin{figure} 
	\includegraphics[angle=-90, width=0.9\linewidth]{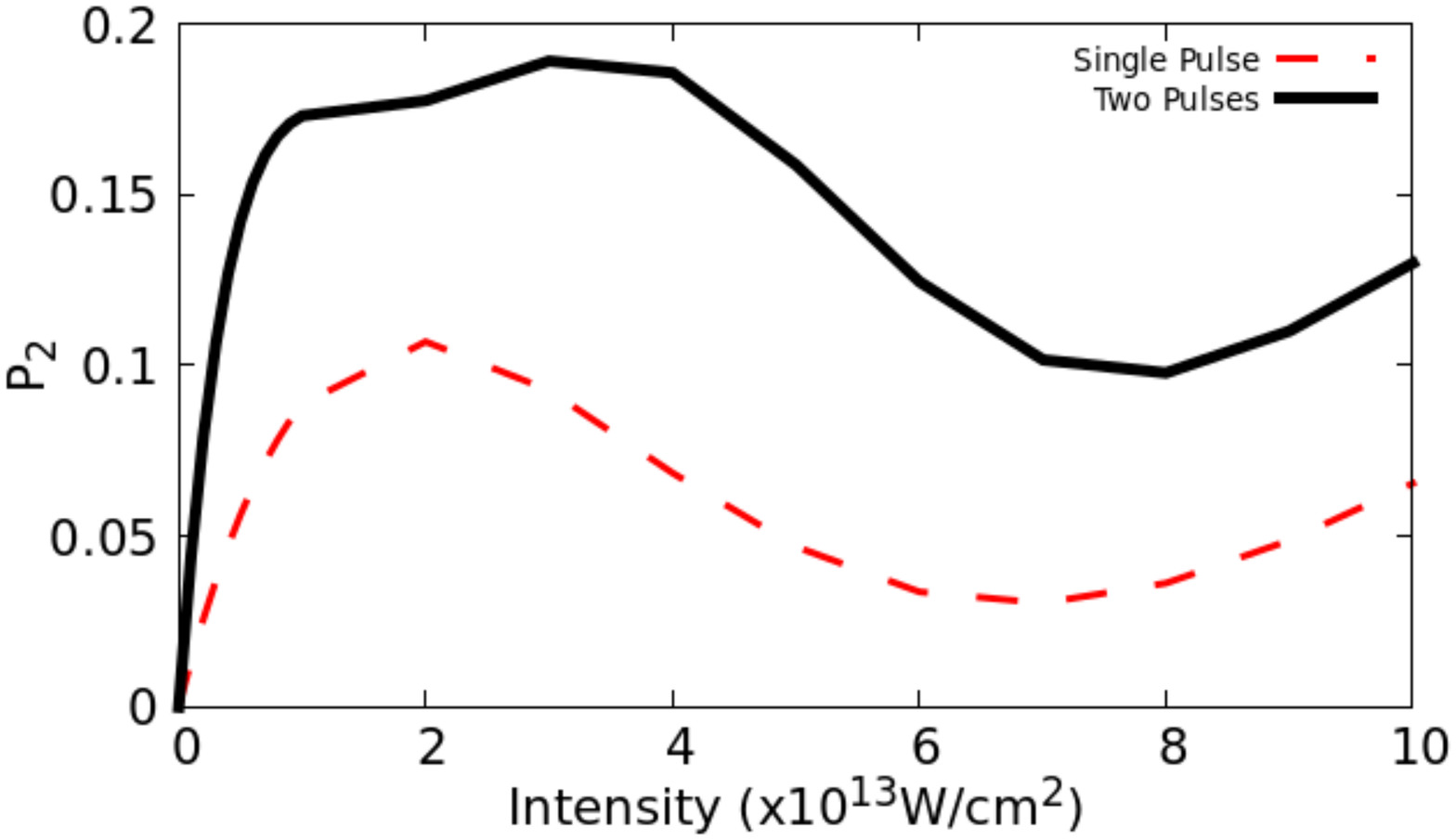}
	\caption{Excited state population $P_2$ as a function of the leading intensity. Red dashed thin line for a single pulse, black thick solid line for two equal fluence pulses, with a delay of $\tau=13fs$.}
	\label{field_regimes}
\end{figure}
Figure \ref{field_regimes} plots the bright donor state population $P_2$ at final time ($t$=50 fs) as a function of the field leading intensity. As is clear from the figure, for the single excitation, the weak field regime extends up to intensities $I=10^{13} W/cm^2$, for which $P_2$ increases rather linearly. The strong field regime shows non-linear behaviors: First a saturation in $S_2$ population, for about $I=2 \times 10^{13} W/cm^2$, followed by a decrease down to $I=7 \times 10^{13} W/cm^2$, corresponding to a partial population trapping in the ground state $S_0$, and then again an increase. We checked that the limit between weak (linear behavior) and strong fields (non-linear behaviors) remains practically unchanged when applying two pulses of equal fluence with a typical delay of $\tau$=13 fs, as will basically be the case for the following control issues.

\section{Results and discussion.}\label{sec4}
We will successively examine the efficiency and the robustness of the interference and kick mechanisms as implemented in their respective control strategies, when going from a reduced 4D to a full 24D model. The guiding principle is to define the parameters of the control field in a 4-dimensional realistic and tractable model. This field once obtained, is later used in a full dimensional dynamic calculation including all the 24 degrees of freedom. The challenge is to discover the efficiency and robustness of such a control field that persists when confronted with the presence of the other numerous degrees of freedom of the molecule.
  \subsection{Interference mechanism.}
\begin{figure} 
  	\includegraphics[angle=-90, width=1.1 \linewidth]{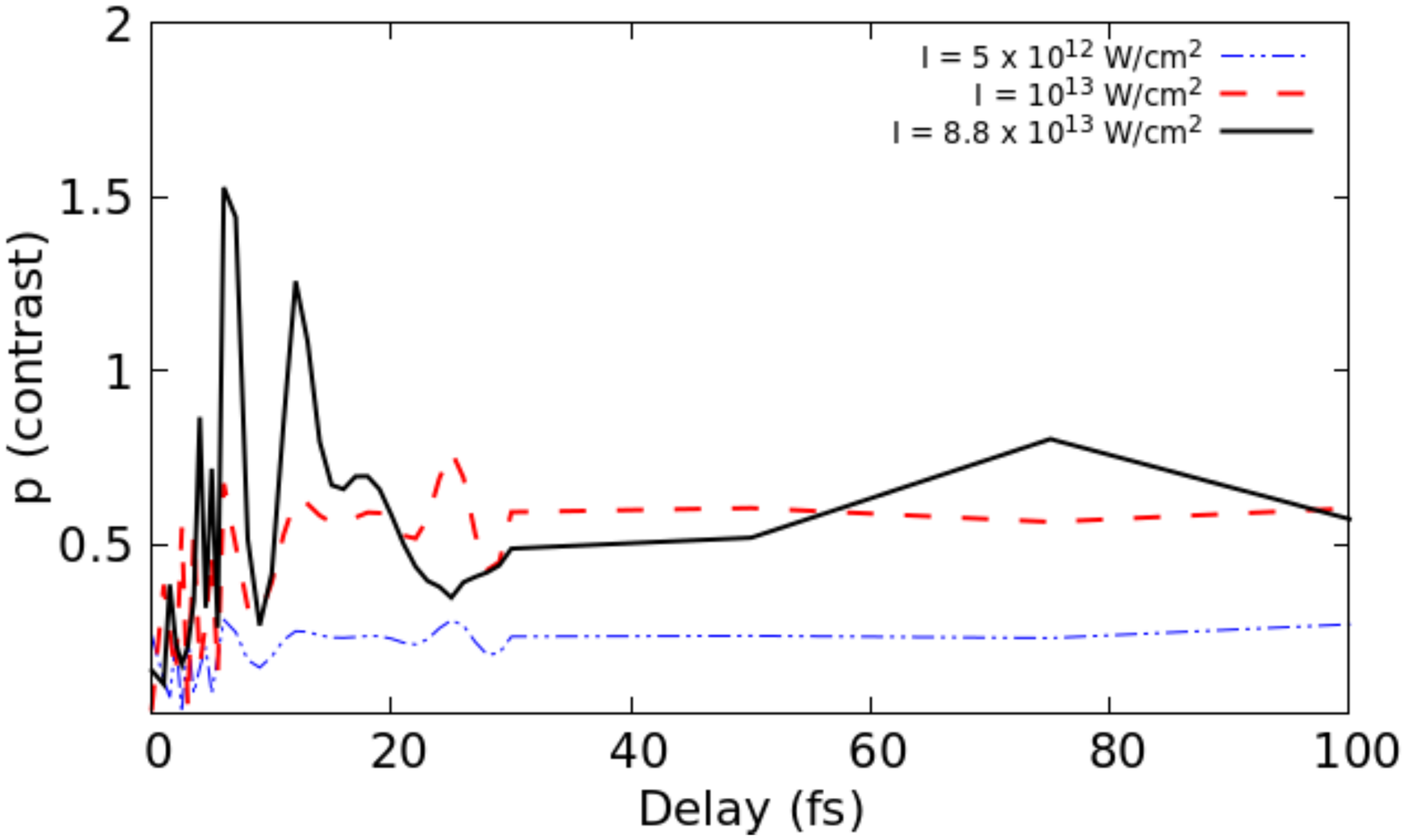}
  	\caption{Asymptotic contrast (4D) as a function of the inter-pulse delay $\tau$ for two pulses, each with a duration of $T=14fs$, equal fluence pulses of leading intensity $I$. The weak field regime is illustrated by the intensities $I=5\times10^{12}W/cm^2$ in blue, dashed-dotted line, and by $I=10^{13}W/cm^2$ in red dashed line.  $I=8.8 \times 10^{13}W/cm^2$ corresponds to the strong field regime, represented by the black solid curve.}
  	\label{p_interference}
  \end{figure}

  Having in mind the general post-pulse evolution of the excited states populations as illustrated by figure \ref{ff_dynamics} two possible interference schemes could be envisioned. One concerns the nuclear wavepacket early dynamics for times typically less than $t<70$ fs within the first vibrational periods, the second corresponding to longer times, typically $t>70$ fs when the wavepacket revisits the FC region, following the revival patterns that are observed in figure \ref{ff_dynamics}. The asymptotic contrast, as defined by Eq.(\ref{contrast}), is calculated in a 4D model as a function of the delay $\tau$ between the two pulses of fixed duration $T$=14 fs. All calculations are conducted within the appropriate field amplitude renormalization condition (cf. Eq.(\ref{fluence})) leading to the same fluence than the one of the single pulse. The results are displayed in figure \ref{p_interference}.
Following the analysis of figure \ref{field_regimes}, we consider three laser intensities: Two pertaining to the weak field regime $I=5\times10^{12}W/cm^2$ and $I=10^{13}W/cm^2$, and one to the strong field regime  $I=8.8 \times 10^{13}W/cm^2$. Weak fields, apart from being easier to realize experimentally, have the advantage of better supporting the first order perturbation approach of Eq.(\ref{route2}). However, the rather low contrast obtained for $I=5\times10^{12}W/cm^2$ makes us favor the choice of $I=10^{13}W/cm^2$ for a typical illustration of the weak field regime. In this regime, the optimal contrast $p=0.8$ is obtained for a delay of $\tau$=25 fs, which, as discussed before, denotes a rather good efficiency, taking into account that less than $20\%$ of the initial state population is transferred to the donor state $S_2$, as seen from figure \ref{field_regimes}. 
\begin{figure} 
	\includegraphics[angle=-90, width=0.9 \linewidth]{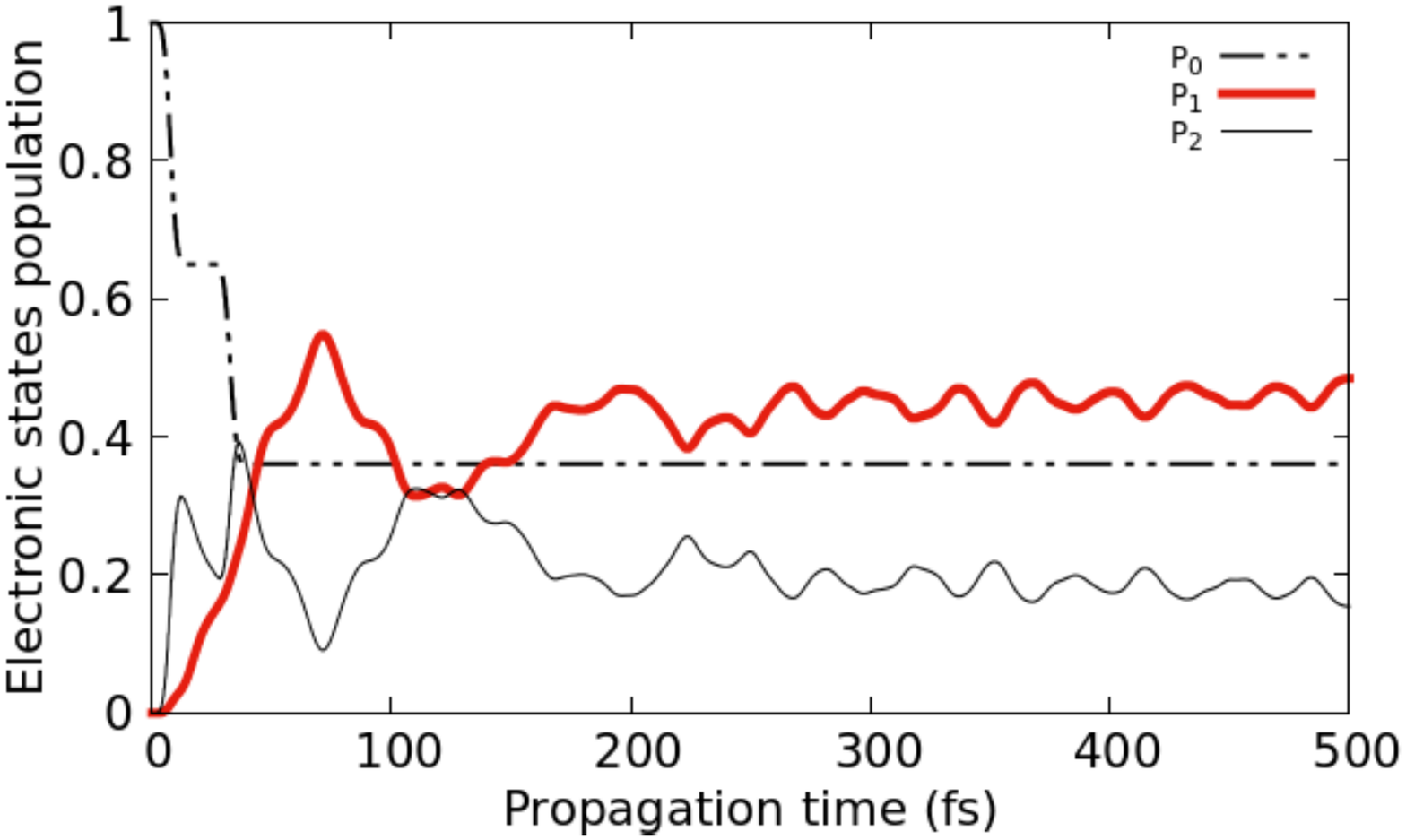}
	\caption{Excited states populations as a function of the propagation time, for two pulses of intensity $I=10^{13}W/cm^2$delayed by $\tau=25fs$. $P_0$ is indicated in dashed-dotted blue line, $P_1$ in thick red solid line, and $P_2$ in thin solid black line.}
	\label{wf2p}
\end{figure}

As a clear indication of the efficiency of the optimal control strategy, we compare the two population transfer dynamics resulting from either a single pulse  or two pulses of equal fluence with the optimal delay of $\tau$=25 fs (figure \ref{wf2p}). The single pulse depletes the ground state up to $P_0=0.65$. The second pulse in constructive interference, produces a depletion up to $P_0=0.38$. The contrast is increased by a factor $\times 3.3$ from $p=0.24$ up to $p=0.8$. The final population which is deposited on state $S_1$ is $P_1=0.5$, meaning that about $50 \%$ of the ground state population is in the acceptor state, while only $16\%$ remains in the donor.

The strong field regime offers much better contrasts by depleting more the ground state population $P_0$. The results displayed in figure \ref{p_interference} show that the maximum is reached for a delay of $\tau$=6 fs and leads to $p=1.55$ as a contrast. Moreover, fast oscillations as a function of the delay show the sensitivity of the observable to this delay, which is another signature of interference effects. Four characteristic delays are considered in the 4D reduced model for an intensity $I=8.8\times 10^{13}W/cm^2$. The single pulse (corresponding to $\tau$=0 fs, after equal fluence renormalization), and $\tau$=9 fs (corresponding to a local minimum in figure \ref{p_interference}) result into comparable dynamics leading to small asymptotic contrasts (typically $p<0.3$). The third choice for the delay $\tau$=6 fs corresponds to the optimal value reached during the early dynamics $p=1.55$, whereas the fourth choice $\tau$=75 fs stands for the optimal value $p=0.9$ obtained during the late dynamics. 
\begin{table}
\setlength{\tabcolsep}{5pt}
\begin{tabular}{ccc}
	\hline
	\hline
	Mechanism & $\omega (eV)$ & $p$ \\
	\hline \\
	Interference & 4.56 & 0.3  \\
	$I = 8.8 \times 10^{13}W/cm^2$ & 4.77 & 1.6 \\
	$\tau$ = 6$fs$ & 5.15 & 0.67 \\
	$T$ = 14$fs$ &  &  \\[1mm]
	\hline \\
	Kicks   & 4.56 & 0.58  \\
	$I = 5 \times 10^{12}W/cm^2$ & 4.77 & 1.71 \\
	$\tau$ = 13$fs$ & 4.91 & 1.64 \\
	$T$ = 10$fs$ &  5.15 & 0.61\\
	\hline
	\hline
\end{tabular}
\caption{Contrasts for several excitation frequencies for both mechanisms discussed in the text. $\omega$ = 4.77$eV$ corresponds to the resonance.}
\label{excitation}
\end{table}
In \emph{a posteriori} way, we also checked the sensitivity and optimality with respect to the resonance condition. The first rows of Table \ref{excitation} for the interference scenario, collect the asymptotic contrasts for two off resonance conditions. The frequency $\omega=4.56 eV$ below the resonance is better fitted for a population transfer towards state $S_1$. But, as such transfers cannot operate in a direct way, the efficiency in terms of contrast, is not satisfactory. The frequency $\omega=5.15 eV$ above the resonance, although opening the possibility for some interesting superposition of vibrational levels  mixing the two excited states, turns out to be less efficient. This study validates the choice made for the resonance frequency, as the presumably optimal one.
\begin{figure} 
	\includegraphics[angle=-90, width=0.9 \linewidth]{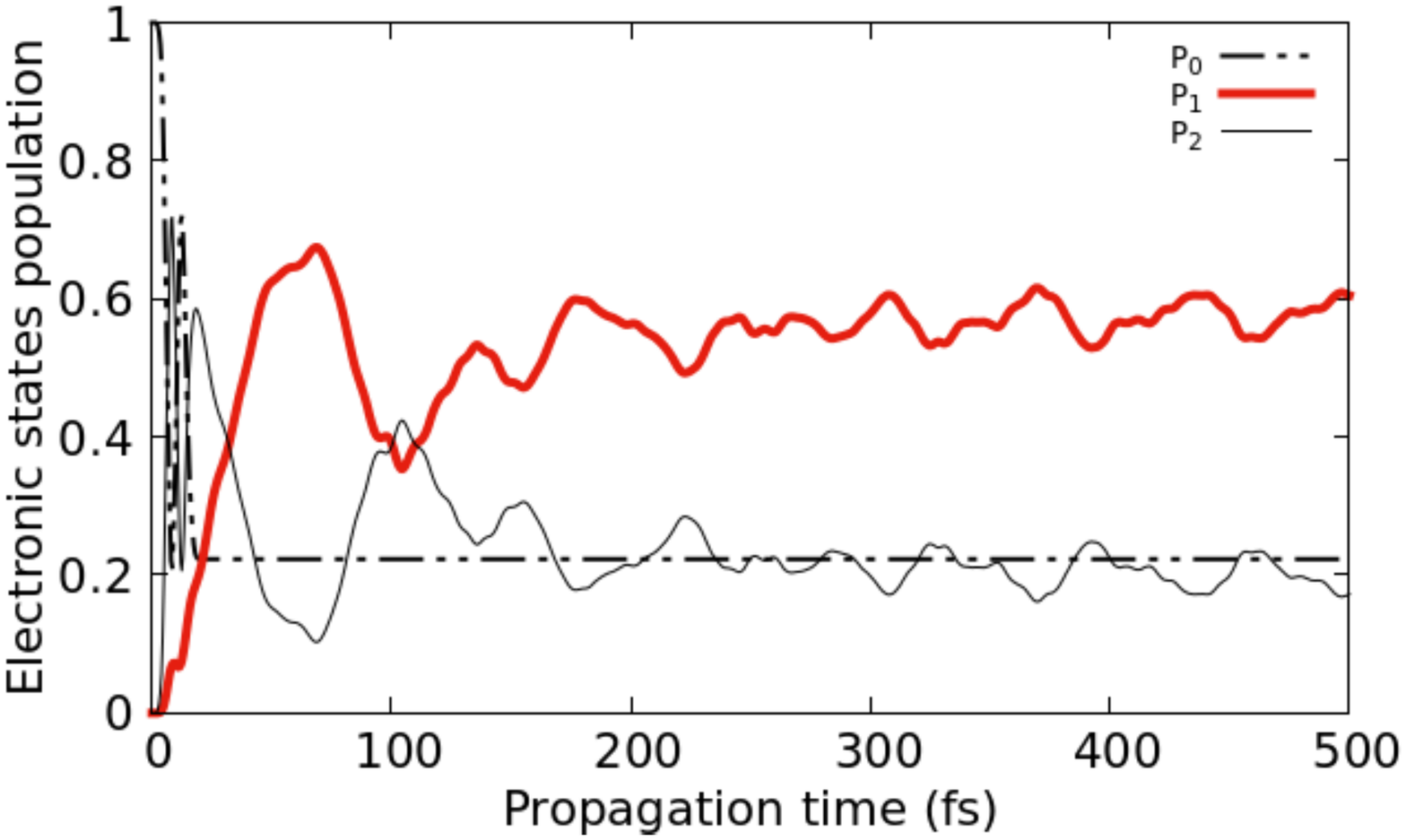}
	\caption{Excited states populations as a function of the propagation time, for two pulses of intensity $I=8.8\times10^{13}W/cm^2$ delayed by $\tau=6fs$ from a full 24D model. $P_0$ is indicated in dashed-dotted blue line, $P_1$ in thick ed solid line, and $P_2$ in thin solid black line.}
	\label{24D_6fs}
\end{figure}
\begin{figure} 
	\includegraphics[angle=-90, width=0.9 \linewidth]{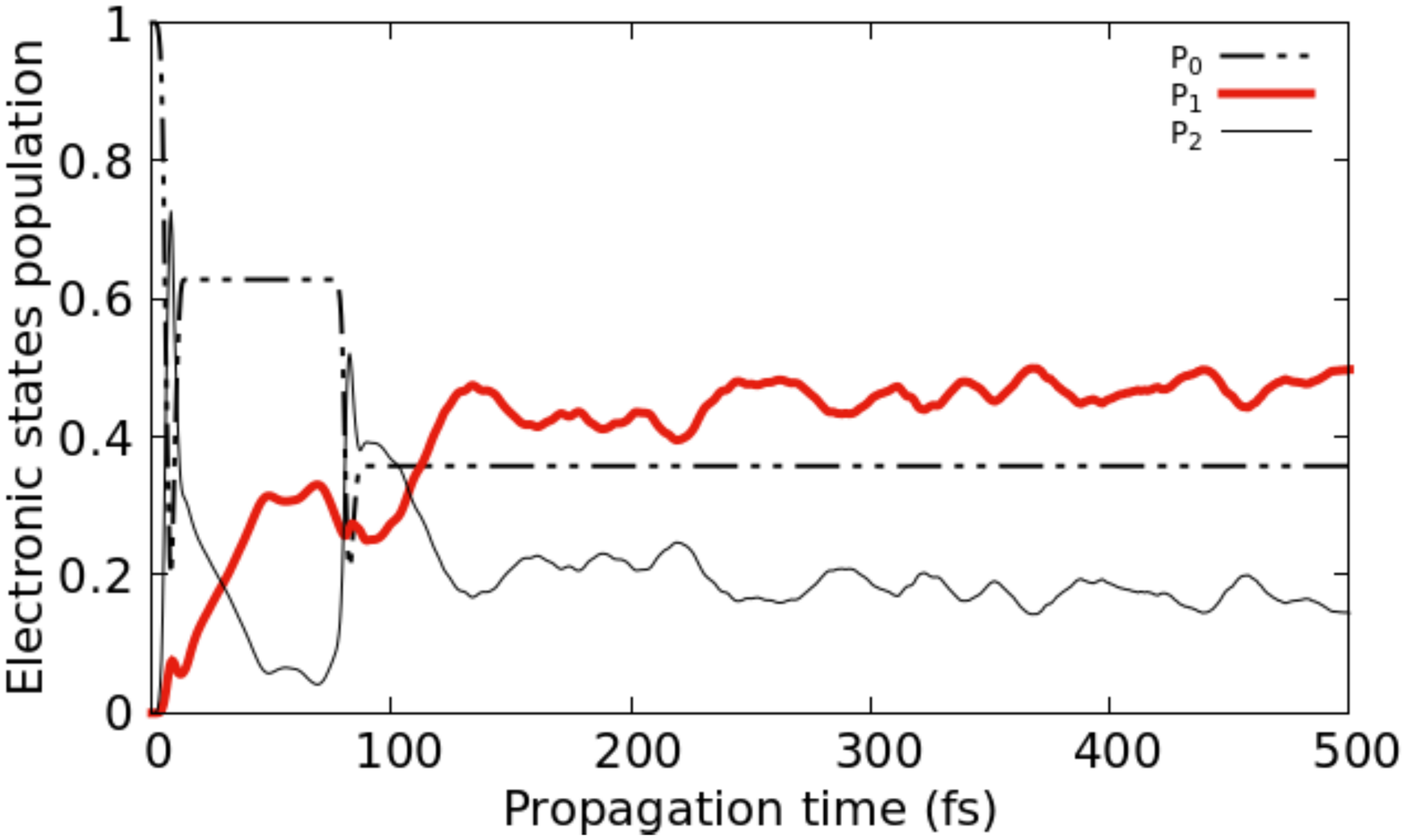}
	\caption{Excited states populations as a function of the propagation time, for two pulses of intensity $I=8.8\times10^{13}W/cm^2$ delayed by $\tau=75fs$ from a full 24D model. $P_0$ is indicated in dashed-dotted blue line, $P_1$ in thick red solid line, and $P_2$ in thin solid black line.}
	\label{24D_75fs}
\end{figure}

Finally, we introduce the optimal control parameter values obtained from the reduced 4D dynamics into a full 24D model describing populations evolution. The results are displayed in figure \ref{24D_6fs} for the early dynamics and figure \ref{24D_75fs} for the late dynamics. It is worthwhile noting that such full dimensional calculations are done once the laser parameters are optimized within the context of the reduced 4D model. As for the 24D  calculations, they necessitate a Central Processing Unit (CPU) time of about 32 hours using OpenMP parallelization scheme  
with 16 processors on a Intel E5-2665 computer. The most striking observation is that the population evolution in the 24D full dynamics closely follows the one of the 4D calculations. Both end with very comparable asymptotic contrast exceeding $p\simeq1.6$ for $\tau$=6 fs and $p\simeq 0.9$ for $\tau$=75 fs. Following this observation, late dynamics control turns out to be less efficient. Actually, better interference schemes are operating for early dynamics where the vibrational wavepacket dynamics is more accurately periodic, with better marked revival structures, better adapted for the interference mechanism. The asymptotic contrasts obtained in the 24D model are comparable with the ones of the reduced 4D model, and even very slightly better. This is presumably due to the fact that the wavepacket is more efficiently and rapidly dispatching towards the additional degrees of freedom, resulting into a faster population stabilization between the excited states. As a conclusion, the optimal result obtained for an early dynamics control scheme, gives an acceptor dark state population of about $P_1=0.6$, which means that $60\%$ of the ground state initial population is transferred to the acceptor $S_1$, while only $16\%$ remains on the donor $S_2$.

 \subsection{Kick mechanism.}
\begin{figure} 
 	\includegraphics[angle=0, width=0.9 \linewidth]{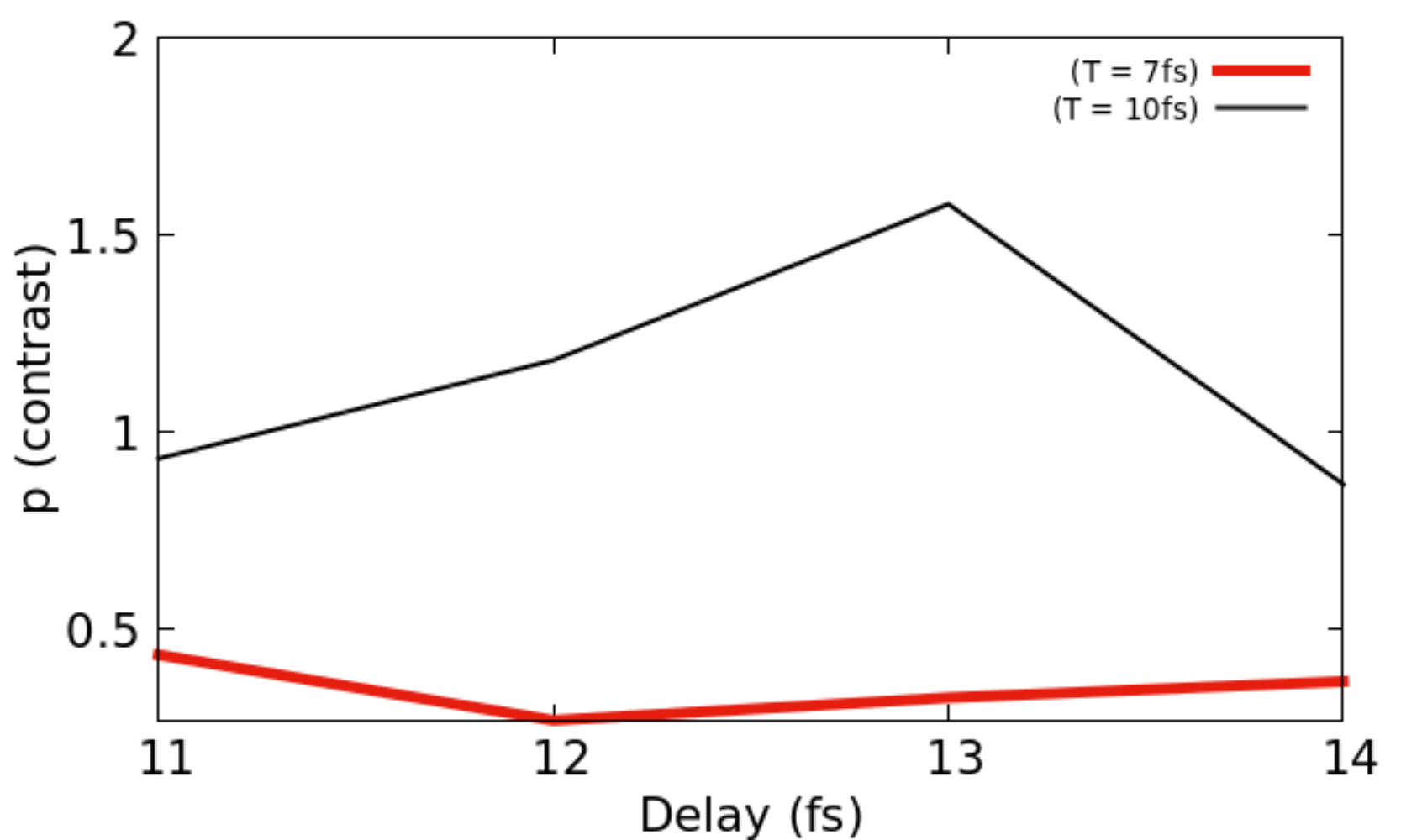}
 	\caption{Asymptotic contrast (4D) as a function of the inter-pulse delay $\tau$ for five equal kicks at a fixed intensity $I=5\times10^{12}W/cm^2$ . The thick black solid line is for kicks duration $T=10fs$; thin red solid line for $T=7fs$.}
 	\label{p_kick}
 \end{figure}

 The kick mechanism operates with ultra-short (broad band) pulses, leading to successive sudden, non-resonant momentum transfer to the molecule \cite{Sugny}. Figure \ref{p_kick} displays the asymptotic contrast as a function of a constant single delay $\tau$ between the kicks imparted to the system from a train of five identical pulses, with fixed intensity $I=5\times10^{12} W/cm^2$. Here, we deal with two control parameters, namely the pulse duration $T$ and the delay $\tau$. It is important to note that a precise value for the excitation frequency is not very relevant, since the broad-band character of the individual pulses renders the strategy essentially non-resonant.
As with the previously discussed mechanism, once the optimal pulse parameters are determined on the reduced more tractable 4D model, they are transposed to the full 24D dynamics. More precisely, we obtain from figure \ref{p_kick}, an optimal asymptotic contrast $p=1.71$, for individual pulse duration $T$=10 fs and delay $\tau$= 13 fs. 
Here also we checked the sensitivity of the results with respect to the excitation frequency $\omega$. The last rows of Table \ref{excitation} corresponding to the kick scenario, collect the results for three off-resonant frequencies. As expected, the contrast is not very sensitive to frequencies within a reasonable window covering $\omega=4.77eV$ to $\omega=4.91eV$ leading to almost the same $p=1.7$. It is only far from these values that the results differ but with a less efficient control.

\begin{figure} 
 	\includegraphics[angle=-90, width=0.9 \linewidth]{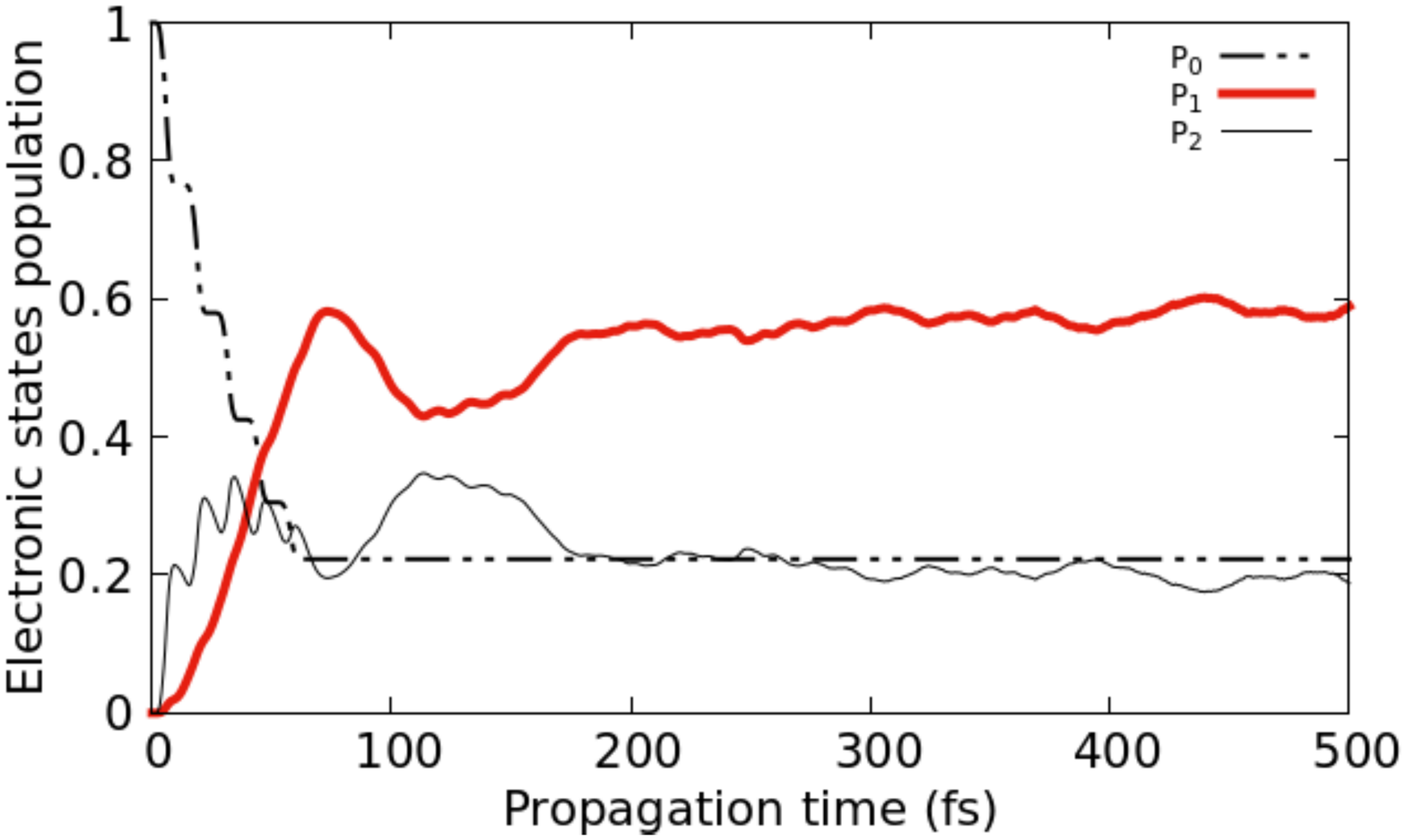}
 	\caption{Excited states populations as a function of the propagation time, for five ultra-short $T=10fs$ pulses of intensity $I=5\times10^{12}W/cm^2$ delayed by $\tau=13fs$ for a full 24D model. $P_0$ is indicated in dashed-dotted blue line, $P_1$ in thick red solid line, and $P_2$ in thin solid black line.}
 	\label{24D_3fs}
 \end{figure}

Figure \ref{24D_3fs} displays the corresponding time evolution of the electronic states populations. 
As can be seen from the step-by-step decreasing evolution of $P_0$, at each kick , the momentum  imparted to the system induces an increasing population towards the excited states. Our  primary aim was to increase regularly the population $P_1$. Such a control could presumably be possible with different delays between the kicks. But for easier experimental requirements, we assume in this control scheme, a  maximum of equally delayed, five identical pulses pertaining to a weak field regime $I=5\times10^{12}W/cm^2$. With such restrictions, the population $P_1$ is not regularly increasing, even during the time when the kicks are applied. Another strategy could be to apply kicks every time $P_1$ starts to decrease, as has been previously done in ref.\cite{Sugny}. Targeting experimental feasibility, within this restrictive parameters sampling space, we get very encouraging results, with contrasts up to $p=1.75$. We also observe from the full dimensional calculation of figure \ref{24D_3fs} that $60\%$ of the population of the ground state  is transferred to the dark acceptor, while only $16\%$ remains on the donor. About 54 CPU hours are needed for this calculation with the same computer as mentioned before.
It is worthwhile noting that short duration and low intensity five-pulses trains can do as well, and even better than the two interfering long duration (resonant) intense pulses, as can be observed from figures \ref{24D_3fs} and \ref{24D_6fs}.

\section{Conclusion.}\label{sec5}
In this work, we consider the intense laser control of funneling dynamics between electronic states with decreasing energy gradient and passing through a conical intersection. Pyrazine molecule offers such a framework involving two excited electronic states, presenting a strong non-adiabatic coupling in relation with their conical intersection. Moreover, the initial ground state is radiatively coupled to the highest energy excited state, which is considered as a bright donor (D). We are concerned by the population transfer dynamics from this state, to the lowest energy dark acceptor (A) state, motivated by the possibility of long term deposit in (A), like in light-harvesting systems. We propose some efficient and robust external control schemes to achieve stable optimal final population in (A). Facing an indirect population transfer process, our control observable should incorporate both a maximum depletion of the ground state and a maximum final population in (A), when sharing between (D) and (A). This is done by defining an asymptotic contrast in terms of the ratio of the population difference between (A) and (D), to the remaining ground state population. In order to maximize this contrast, and inspired by a thorough understanding of the 2D (CI branching space) post-pulse field-free population evolution, we refer to two basic coherent control mechanisms, namely, pump-pump interference and kicks. The search for the optimal control parameters for an electromagnetic field in terms of a train of ultra-short laser pluses is carried on a reduced but still realistic 4D model of pyrazine. More precisely, we optimize the pulse leading frequency and intensity, duration, and the inter-pulse delay. Once obtained and fixed, these parameters are used in the full 24D dynamics to calculate the time evolution of the electronic states populations. It is very gratifying to see that this extension of control mechanisms, often established for over simplified systems involving but a small number of levels, can survive and be effective when applied to systems with a large number of degrees of freedom. This is probably the most relevant message of this study. 

Having in mind the limitations of realistically achievable experimental conditions, the excited states population contrasts we obtained can be considered as efficient ones for pyrazine CI-mediated funneling dynamics. Referring to either interference or kick mechanisms, about $60\%$ of the ground state population is deposited in the acceptor state, while about $16\%$ remains in the donor state. As perspectives, we can mention other control strategies that could be checked in terms of their possible extension to multidimensional systems. One is the strong field pump-probe interference we already mentioned (cf. Eq.(\ref{route1})); the other being the frequency chirp. The density of levels for the vibrational baths being important, addressing in an optimal way some specific levels playing a major role in the funneling dynamics control may deserve an interest.
Even more importantly, pyrazine should be considered but as an example of a modestly complex system. As a conclusion, and consequence of our findings, we anticipate the possibility of potential transposition of such control mechanisms to other larger biological systems. This would certainly be conditioned by having a small number of active normal modes among all the others which are rather spectators in the dynamics we wish to control. Then, similar to what we have observed for pyrazine, external control fields once defined and fixed in low dimensional models, should be effective at higher dimensions.  

\begin{acknowledgments}
	We acknowledge Professors Eric Charron and Michèle Desouter-Lecomte, for fruitful discussions. For computational facilities, we are  indebted to Mesolum of the LUMAT research federation (FR LUMAT 2764).
\end{acknowledgments}

\section*{Data Availability}

The data that support the findings of this study are available from the corresponding author
upon reasonable request.


\bibliographystyle{apsrev}

\end{document}